\documentclass[12pt]{article}
\usepackage{epsfig}

\def\beq{\begin{equation}}
\def\eeq#1{\label{#1}\end{equation}}
\def\eeqn{\end{equation}}

\def\beqa{\begin{eqnarray}}
\def\eeqa#1{\label{#1}\end{eqnarray}}
\def\eeqan{\end{eqnarray}}

\let\bar=\overbar

\def\Dslash{\not{\hbox{\kern-4pt $D$}}}
\def\dslash{\not{\hbox{\kern-2pt $\del$}}}

\def\msb{{\bar{\ssstyle M \kern -1pt S}}}

\def\Title#1{\begin{center} {\Large {\bf #1} } \end{center}}

\long\def\symbolfootnote[#1]#2{\begingroup%
\def\thefootnote{\fnsymbol{footnote}}\footnote[#1]{#2}\endgroup}

\begin{document}

\Title{Lattice QCD in Background Fields%
\symbolfootnote[2]{Talk given by B.~C.~Tiburzi at the Tenth Workshop on Non-Perturbative QCD at l'Institut
d'Astrophysique de Paris, France, 8--12 June 2009.}
}

\bigskip\bigskip


\begin{raggedright}

{\it 
W.~Detmold\index{Detmold, W.},${}^{1,2}$
B.~C.~Tiburzi\index{Tiburzi, B.},${}^{3}$
and
A.~Walker-Loud\index{Walker-Loud, A.}$\, {}^{1}$}
\\
\bigskip
\bigskip
${}^{1}$
{\it Department of Physics, College of William and Mary \\
$\phantom{si}$Williamsburg, VA 23187, USA}
\\
\smallskip
${}^{2}$
{\it
Thomas Jefferson National Accelerator Facility\\
$\phantom{si}$Newport News, VA 23606, USA}
\\
\smallskip
${}^{3}$
{\it Maryland Center for Fundamental Physics\\
$\phantom{si}$Department of Physics, University of Maryland\\
$\phantom{si}$College Park, MD 20742, USA}
\bigskip\bigskip
\end{raggedright}

\begin{abstract}
Electromagnetic properties of hadrons can be computed 
by lattice simulations of QCD in background fields.  We
demonstrate new techniques for the investigation of 
charged hadron properties in electric fields. Our current 
calculations employ large electric fields, motivating us to 
analyze chiral dynamics in strong QED backgrounds, 
and subsequently uncover surprising non-perturbative 
effects present at finite volume. 
\end{abstract}

\section{Introduction and Motivation}

Electromagnetic properties paint an intuitive 
picture of the distribution of charge and magnetism
within a hadron. 
Electromagnetic polarizabilities, 
for example, 
encode the ability of a hadron to deform in response to an applied field. 
This stiffness described by polarizaiblities
is a fundamental property of the hadron, 
with a transparent physical meaning. 
Chiral dynamics, 
moreover,
places tight constraints on the polarizabilities, 
which, in some cases,  
have not compared well with experiment.

Electromagnetic properties of hadrons can be computed 
by lattice simulations of QCD\index{lattice QCD} in background fields\index{background fields}. 
For electromagnetic polarizabilities\index{polarizabilities}, 
the background field method is currently the only option
because computations using four-point functions are not feasible. 
Our lattice QCD computations in background fields and related theoretical developments are summarized here. 
We demonstrate new techniques for the investigation of 
charged hadron properties in background electric fields. Our current 
calculations employ large electric fields, motivating us to 
analyze chiral dynamics in strong QED\index{strong QED} backgrounds, 
and subsequently uncover surprising non-perturbative 
effects present at finite volume.

\section{Background Field Method}

There are two basic steps involved in background field computations of hadron properties. 
First one computes hadron correlation functions using lattice QCD in classical external fields, 
for various values of the external field strength. 
One then compares the calculated correlation functions with their behavior predicted
by the relevant single-particle effective action. 
The parameters of the single-particle effective action can be deduced by considering 
the field-strength dependence of the correlation functions.

The simplest application of the method is for a neutral scalar particle. 
We use the neutral pion to exemplify this. Imagine correlation functions for 
the neutral pion have been calculated in a background electric field
specified by the gauge potential
$A_\mu (x) = - \mathcal{E} x_4 \delta_{\mu, 3}$.
As a composite scalar, 
the most general effective action for the neutral pion in an external electric field 
has the form
\begin{equation} \label{eq:pizero}
\mathcal{L} (\vec{p} = 0)
=
\frac{1}{2} \pi^0
\left[
- \partial_4 \partial_4 
+ m_{\pi^0}^2
+ m_{\pi^0} \, \alpha_E \, \mathcal{E}^2
\right]
\pi^0
.\end{equation}
Here we work in Euclidean space, and have projected the Lagrangian onto vanishing spatial momentum, 
which is possible due to our implementation of the electric field. 
Terms involving higher powers of the electric field are possible, 
but we have kept only the lowest-order contribution as we imagine
the fields are sufficiently weak. 
The coefficient of the second order term, $\alpha_E$, is the electric polarizability.

Using Eq.~(\ref{eq:pizero}), 
one can deduce the behavior of the neutral pion two-point correlation function, 
namely
\begin{equation} \label{eq:neutralpion}
G(\tau) 
= 
\sum_{\vec{x}}
\langle 0 |
\pi^0(\vec{x}, \tau)
\pi^0(\vec{0}, 0)
| 0 \rangle
= 
Z 
\exp 
\left(
- E \tau
\right)
,\end{equation}
where the neutral pion energy, $E$, is given by
$E = m_{\pi^0} + \frac{1}{2} \alpha_E \, \mathcal{E}^2$,
and the sign arises from our treatment in Euclidean space. 
Thus by determining the neutral pion energy as a function
of the applied electric field, one can deduce the electric polarizability.

\begin{figure}[htb]
\begin{center}
\epsfig{file=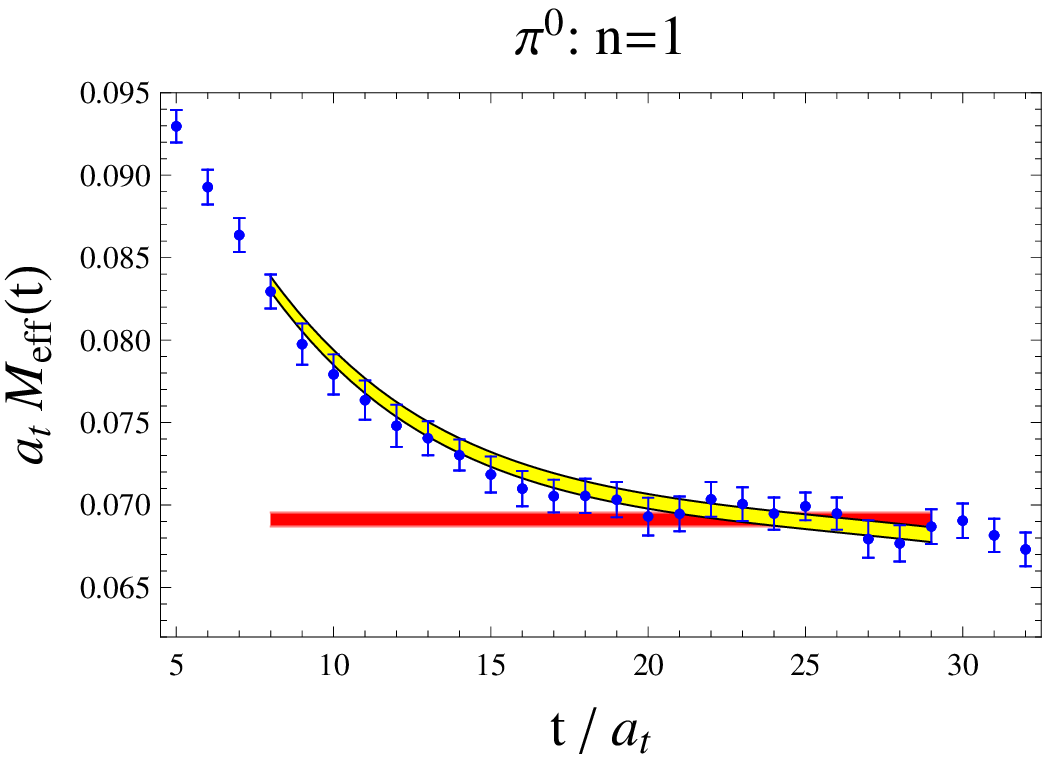,height=1.9in}
\epsfig{file=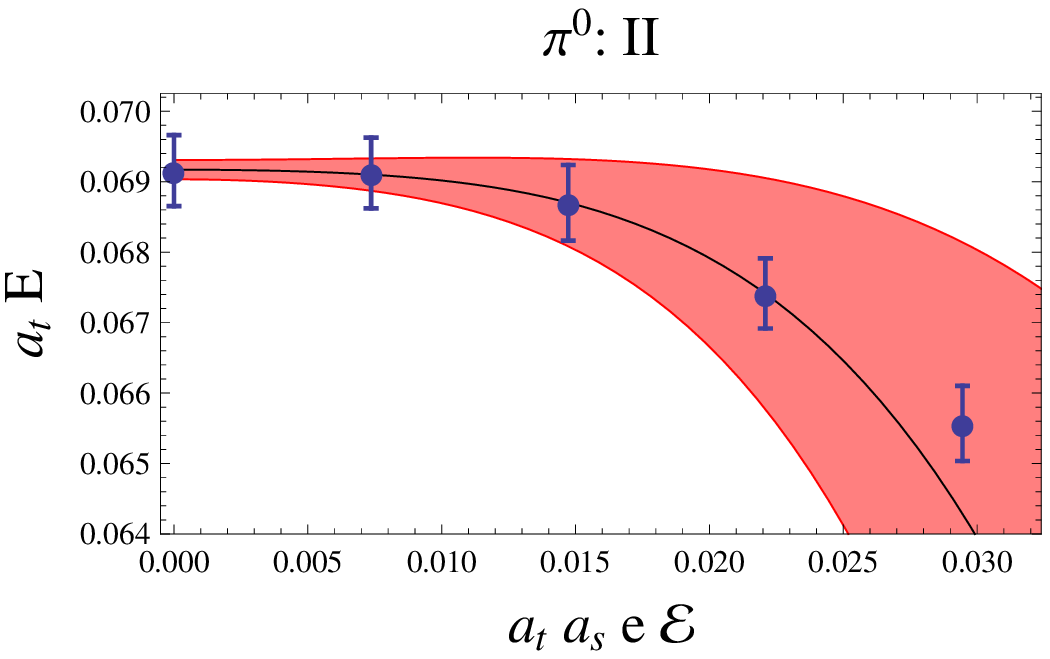,height=1.9in}
\caption{On the left: fit to the neutral pion correlation function in the electric field with strength corresponding to $n=1$. 
Plotted is the effective mass $M_{eff} (t) = - \log G(t+1)/G(t)$.
On the right: a fit (denoted II) to the electric-field dependence of the neutral pion energy.}
\label{f:neutralpion}
\end{center}
\end{figure}

Recently we have carried out this procedure for the neutral pion (additionally for the neutral kaon)~\cite{Detmold:2009dx}. 
We used an ensemble of anisotropic $(2+1)$-flavor dynamical clover lattices generated by the \emph{Hadron Spectrum} collaboration~\cite{Edwards:2008ja,Lin:2008pr}.
Specifically our ensemble consists of $200$ configurations of size $20^3 \times 128$ with light quark masses giving
rise to a pion mass of $390 \, \texttt{MeV}$.  
The anisotropy factor has been tuned to $a_s / a_t = 3.5$. 
We computed neutral pion correlation functions for various values of the electric field, 
and then fit to the expected form in Eq.~(\ref{eq:neutralpion}).
Due to our choice of lattices, we use a two-state fit in order to stabilize the extraction of the ground-state energy. 
In Figure~\ref{f:neutralpion}, 
we show one such fit for the lowest value of the electric field (denoted by $n=1$). 
Additionally
we show the energies extracted as a function of the external field strength. 
For sufficiently small fields, 
the behavior of the energies should be quadratic with respect to the field. 
The fit shown (denoted II) includes both quadratic and quartic terms.


\section{Field Strengths and Strong Fields}

Having investigated the neutral pion energy as a function of the external electric field strength, 
we notice that the values of the external field employed are not necessarily perturbatively small. 
We are not able, however, to simulate at arbitrary values of the field. 
The implementation of a uniform field coupled to matter fields is restricted by 
quantization conditions relevant for a torus~\cite{'tHooft:1979uj}.

For our background electric field, 
the field strength $\mathcal{E}$ must be quantized in the form
\begin{equation}
\mathcal{E} = \frac{2 \pi n}{ q_d L T}
,\end{equation}
where $q_d = - 1/3 \,  e$ is the electric charge of the down quark, 
$L$ is the length of the lattice in the $x_3$-direction, 
$T$ is the length in the $x_4$-direction, 
and $n$ is an integer. 
To implement a uniform field on a torus, 
one must use link variables. 
The standard links, 
$U^{cl}_\mu (x) = \exp [i q A_\mu(x) ] = \exp ( - i q \mathcal{E} x_4 \delta_{\mu, 3}) $,
are not enough to ensure a uniform field on a discrete torus. 
Plaquettes that wrap around the time-boundary see a dramatic
spike in the electric field. 
This effect can be removed by quantizing the fields, and adding 
additional transverse gauge links
$U^\perp_\mu(x) = \exp ( i q  \mathcal{E} T x_3 \delta_{\mu, 4} \delta_{x_4, T-1})$~\cite{Smit:1986fn}.
The net result is to modify the color links $U_\mu(x)$ in the following way
\begin{equation} \label{eq:quantize}
U_\mu(x) \longrightarrow U_\mu(x) U^{cl}_\mu(x) U^\perp_\mu(x)
.\end{equation}
Notice this post-multiplication leaves the sea quarks electrically neutral. 
Charging the sea would require at least an order of magnitude greater computing resources.

\begin{figure}[htb]
\begin{center}
\epsfig{file=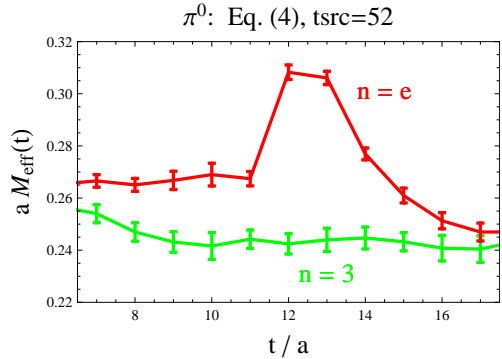,height=1.9in}
\caption{Comparison of quantized and non-quantized field strengths.
In the plot, 
correlation functions have been translated forward by twelve units in time.}
\label{f:quantized}
\end{center}
\end{figure}

To investigate the necessity of quantized field strengths, 
we plot in Figure~\ref{f:quantized} the neutral pion correlation function
calculated in two different classical fields. 
Both fields are implemented using Eq.~(\ref{eq:quantize}), 
but we use the values
$n = e = 2.71828\ldots$, 
and 
$n = 3$. 
While numerically the fields are comparable in magnitude, 
the non-integer value results in a single plaquette at the far corner 
of the lattice with a dramatic non-uniformity. 
We can see the effect of the non-uniformity by letting the 
neutral pion propagate around the time boundary.
For this exploration, 
we use configurations with $64$ sites in the time direction,
and place the pion source at $t = 52$
(details of this study were presented in~\cite{Detmold:2008xk}). 
After translating the correlation functions forward by twelve units, 
the onset of a plateau in the effective mass at around
$t = 12$ corresponds to 
the edge of the lattice in the time direction. 
A plateau is only seen for the $n=3$ value of the field.

While current lattice sizes give rise to rather large values of the external fields, 
we are still in the regime of a so-called strong field power counting in chiral perturbation theory. 
On the $20^3 \times 128$ anisotropic lattices, 
we have 
$(e \, \mathcal{E}  / m_\pi^2 )^2 = 0.18 n^2$, 
where $n$ is the integer in the field strength quantization condition. 
For such values of the field, 
we can perform chiral perturbation theory calculations to determine
the response of the pion cloud to the strong field~\cite{Tiburzi:2008ma}. 
Such calculations treat chiral corrections perturbatively
$m_\pi^2 / \Lambda_\chi^2 \ll 1$, 
but re-sum charge couplings to all orders, 
$e \, \mathcal{E} / m_\pi^2 \sim 1$.

\begin{figure}[htb]
\begin{center}
\epsfig{file=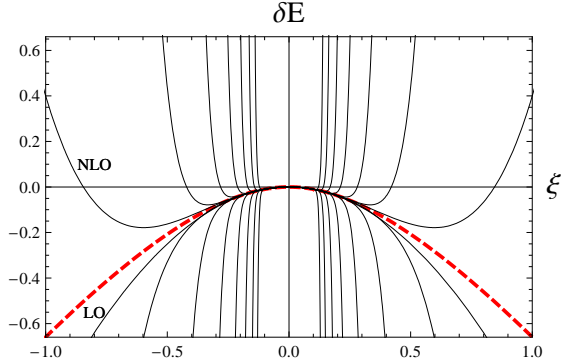,height=1.9in}
\caption{Plot of the electric field dependence of the neutral pion energy. 
The dashed curve shows the non-perturbative result at one-loop order
in the strong field chiral expansion. LO and NLO represent the leading and
next-to-leading order perturbative expansions of the energy in powers of 
the electric field, where LO $\propto \mathcal{E}^2$ and NLO includes terms $\propto \mathcal{E}^4$.
Successive perturbative approximations are also depicted.}
\label{f:pertvsnonpert}
\end{center}
\end{figure}

As an example of a strong field calculation in chiral perturbation theory, 
one can derive the field-strength dependence of the neutral pion energy 
in strong electric fields. 
The shift, $\delta E$, of the neutral pion energy has the form
\begin{equation} \label{eq:nonpert}
\delta E 
= 
\frac{(4 \pi f)^2}{m_\pi^2} 
\left[ 
\frac{E( \vec{p} = 0) - m_\pi}
{m_\pi}
\right]
= 
\frac{e \, \mathcal{E}}{m_\pi^2} \,
\mathcal{I} \left( \frac{m_\pi^2}{e \, \mathcal{E}} \right)
,\end{equation}
where the explicit expression for $\mathcal{I}(x)$ arising at one-loop order
has the form
\begin{equation}
\mathcal{I} (x)  
= 
x \left( 1 - \log \frac{x}{2} \right)
+ 2 \log \Gamma \left( \frac{1 + x}{2} \right)
- \log 2 \pi 
.\end{equation}
In Figure~\ref{f:pertvsnonpert}, 
we plot the neutral pion energy
as a function of the electric field, 
$\xi = e \, \mathcal{E} / \sqrt{6} m_\pi^2$. 
The non-perturbative result of Eq.~(\ref{eq:nonpert})
is contrasted with the first two non-vanishing 
perturbative approximations. 
In the range of $\xi$ shown, 
the perturbative expansion has broken down:
higher order perturbative approximations result
in worse agreement with the non-perturbative result.
Consequently the LO approximation does best.
This behavior is characteristic of asymptotic expansions.

Finally notice that on the same lattices, 
magnetic fields are further from the perturbative regime,
i.e.~we have
$( e \, B / m_\pi^2 )^2 = T^2 / L^2 (e \, \mathcal{E} / m_\pi^2)^2 $,
due to the quantization condition, 
with $T^2 / L^2 = 3.3$ for our anisotropic lattices.

\section{Charged Pion}

The background field method can also be applied to charged particles, 
as was first suggested in~\cite{Detmold:2006vu}. 
The philosophy is the same: 
compute the hadron correlation function in various values of the external field strength, 
and then match onto the behavior expected from the single-particle effective action. 

Let us use the charged pion as an example. 
Imagine we have computed the charged pion correlation function for various
values of an external electric field. 
What should we expect for the behavior of these correlation functions?
For the charged pion, 
the single-particle effective action has the form
\begin{equation} \label{eq:charged}
\mathcal{L} ( \vec{p} = 0) 
= 
\pi^\dagger
\left[
- \partial_4 \partial_4 
+ 
\mathcal{E}^2 x_4^2
+ 
E^2(\mathcal{E}) 
\right]
\pi
.\end{equation}
The explicit time-dependent term arises from minimal coupling of the electric field to the charged pion.
The non-minimal couplings have been subsumed into the rest energy, $E(\mathcal{E})$, 
which has a small-field expansion in powers of the electric field. 
The coefficient of the first field-strength dependent term in this expansion is proportional to the polarizability.

\begin{figure}[htb]
\begin{center}
\epsfig{file=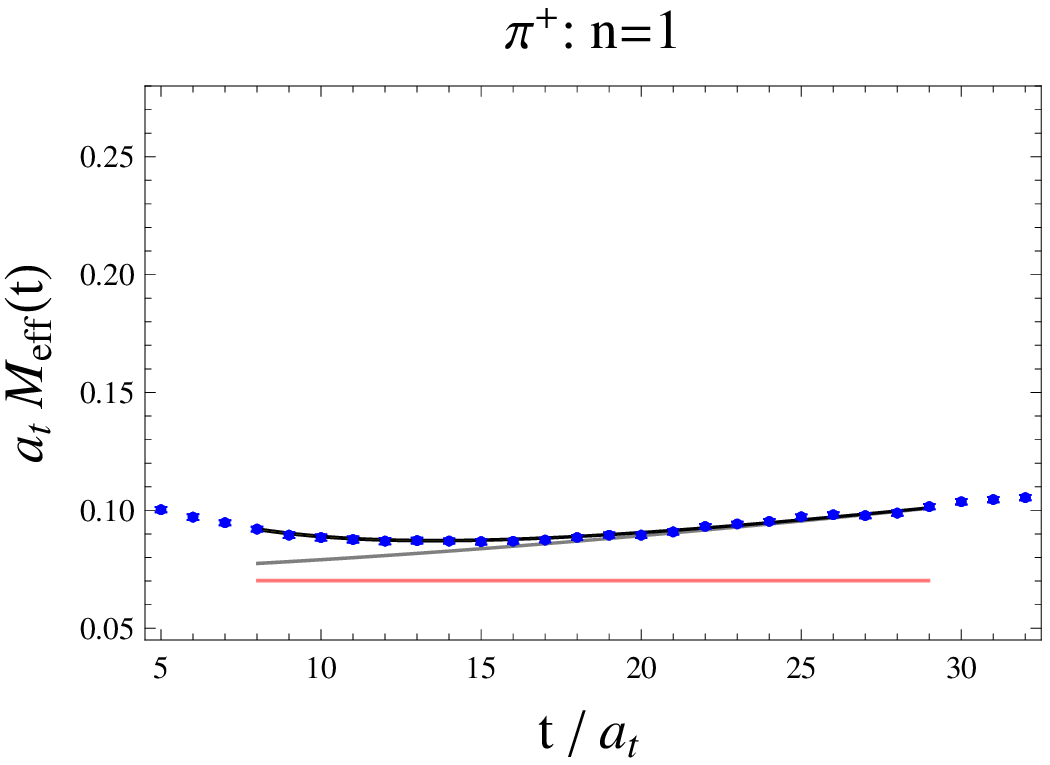,height=1.9in}
\epsfig{file=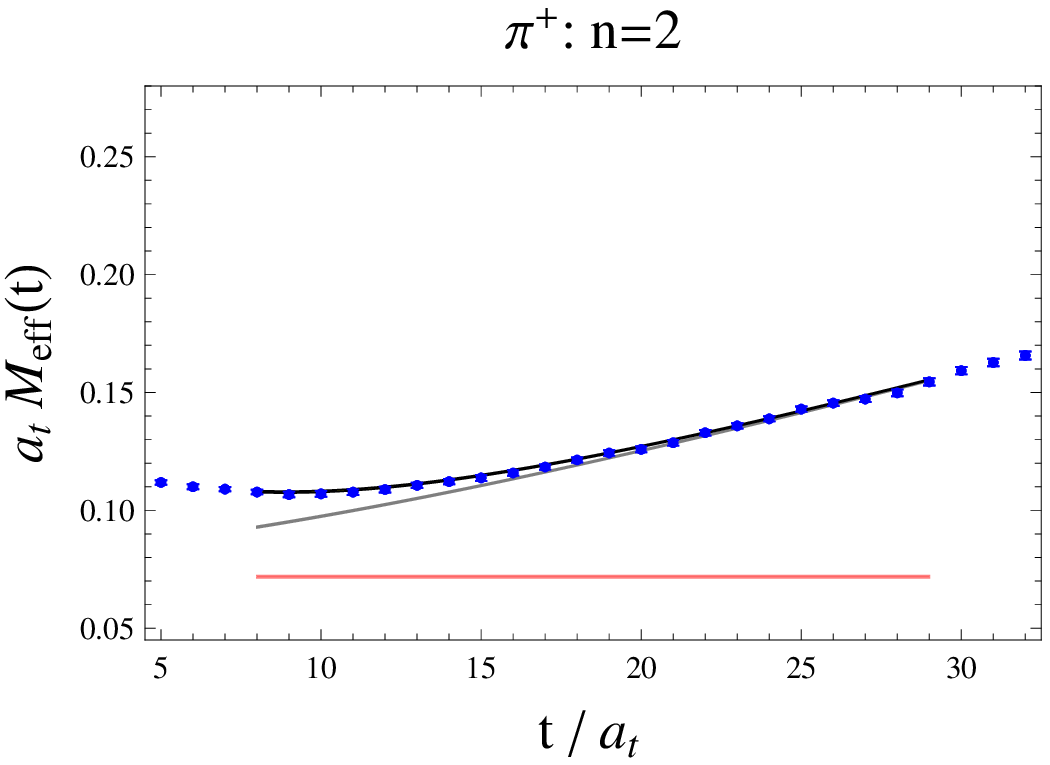,height=1.9in}
\epsfig{file=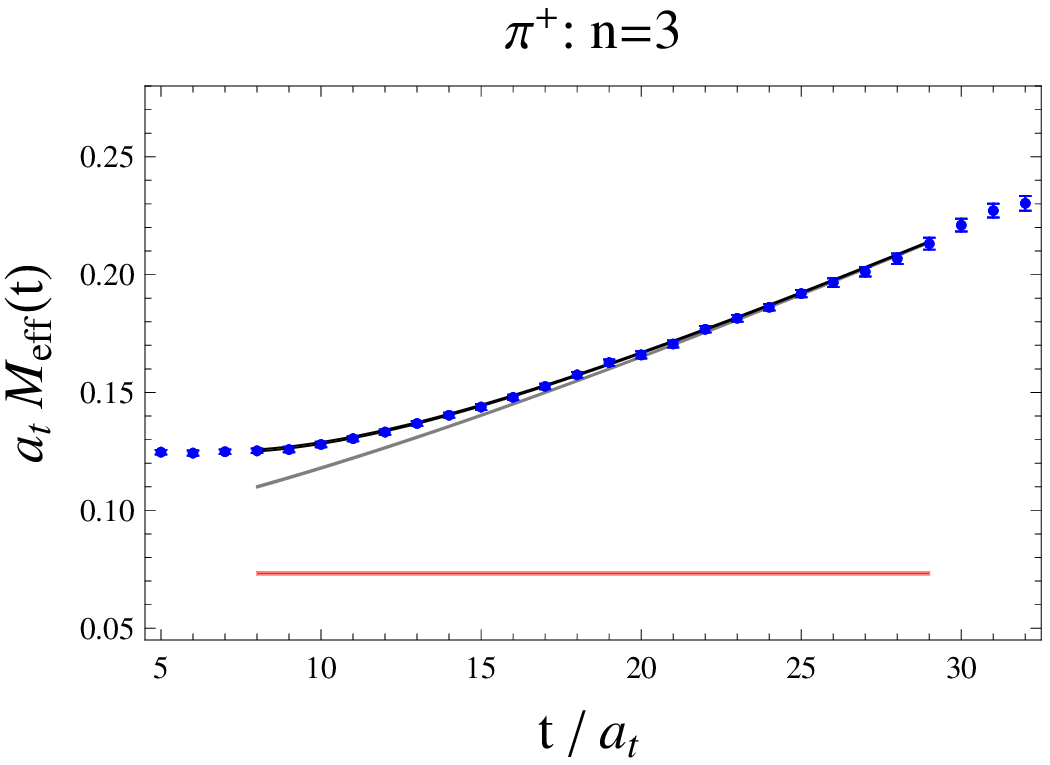,height=1.9in}
\epsfig{file=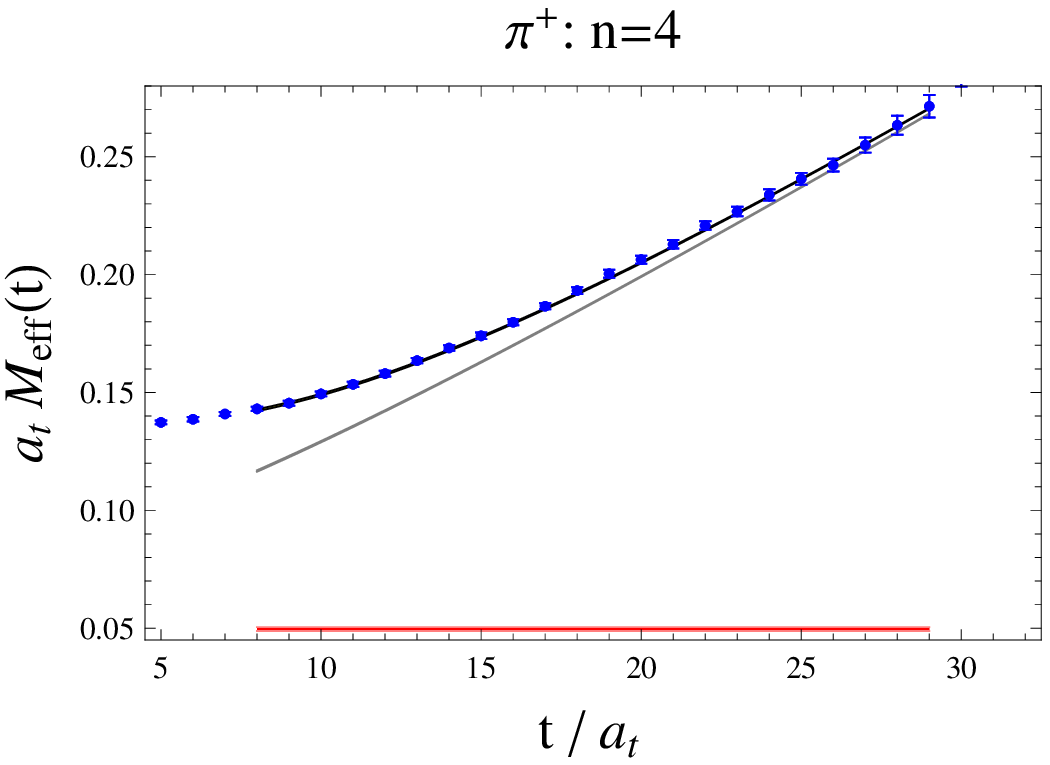,height=1.9in}
\caption{Charged pion correlation functions in external electric fields.}
\label{f:chargedpion}
\end{center}
\end{figure}

Using Eq.~(\ref{eq:charged}), 
we deduce the form of the charged pion correlation function, 
namely
\begin{equation} \label{eq:chargedpion}
G(\tau) 
= 
\sum_{\vec{x}}
\langle 0 |
\pi^+(\vec{x}, \tau)
\pi^-(\vec{0}, 0)
| 0 \rangle
= 
Z \,
\mathcal{D}_{\frac{\mathcal{E} - E^2}{2 \mathcal{E}}} \left( \sqrt{ 2 \mathcal{E} \tau^2 } \right)
,\end{equation}
where $\mathcal{D}_\nu(z)$ is a parabolic cylinder function
(for which a convenient integral representation can be derived using Schwinger's proper-time trick~\cite{Schwinger:1951nm,Tiburzi:2008ma}). 
In Figure~\ref{f:chargedpion}, 
we show the behavior of the charged pion correlation function for four values of the external field 
strength, corresponding to the integers $n = 1$--$4$. 
Shown along with the correlation functions are fits using the relativistic scalar propagator,
Eq.~(\ref{eq:chargedpion}). 
As with the neutral case, we have performed a two-state fit. 
The anisotropy is of considerable help yielding $3.5$ times the number of time slices over which to fit the correlation function. 
While the pion charge leads to correlation functions with dramatically different time-dependence than a simple exponential falloff, 
the behavior seen is inline with expectations for a relativistic charged scalar in a constant field.

\section{Finite Volume Artifacts}

While the lattice volume restricts the strength of external fields, 
there are additionally finite size effects of a dynamical nature. 
Physically one expects rather sizable modification to the electromagnetic polarizabilities
in finite volume because these quantities physically have the dimension of volume.
Furthermore a hadron's deformation arises from its charged pion cloud, 
and lattice volumes are not considerably greater than the pion Compton wavelength,
$\lambda_C = 1 / m_\pi \sim 3/2 \, \texttt{fm}$.
It is na\"ive, 
however, 
to merely ascertain how much of the induced dipole moment owes to finite volume~\cite{Hu:2007ts,Tiburzi:2007ep}.
As multipole polarizabilities arise from $SO(3)$ invariant physics, 
we must instead deal with the subtleties of electromagnetic fields on a torus, 
which only maintains a discrete cubic subgroup of rotations.

A related subtlety in the physics on a Euclidean torus is the lack of boost invariance.
Matrix elements in the rest frame are related to those in a boosted frame only
up to a finite volume modification. 
Thus the rest frame matrix element for the pion charge
\begin{equation}
\langle \pi(\vec{0}\, ) | J_4 | \pi (\vec{0}\,) \rangle 
= 2 p_4 Q
\end{equation}
does not generate the current $\vec{J} = Q \vec{v}$ when boosted to an arbitrary frame, 
\begin{equation}
\langle \pi(\vec{p} \, ) | \vec{J} \, | \pi (\vec{p} \, ) \rangle 
= 2 \vec{p}  \, [ Q - \Delta Q(L) ]
.\end{equation}
This fact is consistent with Ward-Takahashi identities, and is analogous to charge screening at finite temperature~\cite{Hu:2007eb}.
While Ward identities are violated, they are not applicable in finite volume because a limiting 
procedure does not exist to take derivatives with respect to the quantized momentum modes, 
$\vec{p} = 2 \pi \vec{n} / L$. 
One can use coordinate space methods in finite volume to derive
finite volume corrections using chiral perturbation theory~\cite{Tiburzi:2008pa}. 
Specifically one can derive the single-particle action in finite volume; which, for the charged pion,
has the form
\begin{equation}
\Delta \mathcal{L}(\vec{p} = 0) 
= 
\pi^\dagger 
\left[
\Delta Q^2(L) \mathcal{E}^2 x_4^2 + \Delta \alpha(L) \mathcal{E}^2
\right]
\pi
,\end{equation}
where only terms up to $\mathcal{O}(\mathcal{E}^2)$ have been retained.

\begin{figure}[htb]
\begin{center}
\epsfig{file=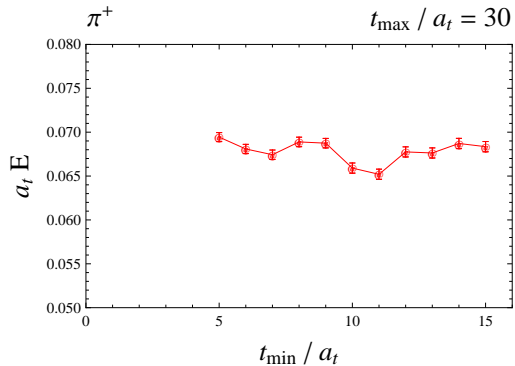,height=1.9in}
\caption{Extracted charged-pion rest energies as a function of the fit window for the $n=3$ field strength. 
The fit window spans $[ t_{min}, t_{max}]$, with $t_{max} = 30$ fixed.}
\label{f:oscillate}
\end{center}
\end{figure}

To our surprise, 
the perturbative expansion of finite volume corrections is not sufficient to account 
for all of the systematics present in the lattice data. 
Shown in Figure~\ref{f:oscillate} is an unexpected systematic effect:
the extracted rest energy of the charged pion shows sensitivity to the 
fit window chosen. 
Finite size effects can give rise to such oscillatory time dependence 
as we now argue.

Using our gauge potential for the electric field, 
there are other gauge invariant quantities on a torus besides
the field strength $\mathcal{E}$. 
These quantities are Wilson loops formed using the gauge holonomy,
$\Theta$, 
given by
\begin{equation}
\Theta = \int_0^L dx_3 \, A_3 (x) = - \mathcal{E} L x_4
.\end{equation}
Thus the Wilson loops, $W (\Theta) = e^{i \Theta}$, will be time dependent, 
and such oscillatory behavior cannot be captured in perturbation theory in the strength of the field. 
One needs a finite volume calculation that treats 
$\Theta(x_4) \sim 1$. 
Results of these chiral perturbation theory calculations are rather cumbersome to work with, 
and we can make considerable progress by appealing to symmetries.

Consider the neutral pion effective action in finite volume. 
The operators appearing are constrained by parity, charge conjugation, gauge invariance, and discrete time translation.
Writing down all terms consistent with these symmetries we find, 
\begin{equation}
\mathcal{L} (\vec{p} = 0) 
= 
\frac{1}{2}
\pi^0
\left[
- \partial_4 \partial_4
+ 
E^2( \mathcal{E})
+ 
\sum_{n = 1}^\infty F_n \,
\mathcal{W}(n \Theta)
\right]
\pi^0
,\end{equation}
where all parameters implicitly depend on the lattice size $L$. 
The relevant Wilson loops, 
$\mathcal{W}(x)$,  
are Hermitian, 
and given by 
$\mathcal{W}(x) = \frac{1}{2} [ W(x) +  W^\dagger(x)]$. 
Physically the tower of Wilson loop couplings arises from 
charged virtual pions propagating around the lattice $n$ times.  
Given that the probability for such propagation 
scales exponentially, 
$\mathcal{P}_n \propto \exp ( - n \, m_\pi L)$, 
it is reasonable to consider only the first term in the tower of 
Wilson loop operators. 
The resulting single-pion effective action
\begin{equation}
\mathcal{L} (\vec{p} = 0) 
\longrightarrow 
\frac{1}{2}
\pi^0
\left[
- \partial_4 \partial_4
+ 
E^2( \mathcal{E})
+ 
F_1 \cos ( \mathcal{E} L x_4 )
\right]
\pi^0
,\end{equation}
leads to a pion propagator that can be expressed in terms of Mathieu functions. 
It seems likely that such oscillatory functions can explain the variation
of the extracted energies with respect to the fit window. 
This work, however, is still in progress.

\section{Summary}

Above we have reviewed lattice QCD computations in background fields. 
We have stressed the two basic features of such calculations;
namely,
the measurement of hadronic correlation functions in external fields, 
and the matching of the expected behavior for these correlation functions
obtained from single-hadron effective actions. 
For ease we only considered pions in background electric fields.

The neutral pion in infinite volume presents the simplest application of
the background field method: 
a field-strength dependent shift of the pion energy is the only modification
to the neutral pion correlation function. 
Measuring this shift allows one to deduce the electric polarizability. 
The charged pion electric polarizability can also be deduced by using the background field method. 
For this case, 
however, 
the calculated correlation function has non-standard behavior as a function of time. 
Nevertheless, 
this behavior of the correlation function is precisely that predicted for a relativistic charged particle in a
constant electric field.

Our computations are currently limited by systematic errors. 
As field quantization conditions must be met, 
the relatively small lattice volume employed leads to large values of the electric field. 
As one leaves the regime in which the field-strength dependence is perturbative, 
strong field chiral perturbation theory can be applied. 
In fact, 
chiral perturbation theory gives parameter-free predictions for hadron energies 
at next-to-leading order in the chiral expansion as a function of 
$e \, \mathcal{E} / m_\pi^2$. 
Small lattice volumes lead to another malady: 
finite size effects. 
Single pion effective actions are more complicated in finite volume.  
Quantum fluctuations allow virtual pions to wrap around the lattice 
and encounter background fields that differ topologically. 
Couplings of Wilson loops to pions must be included in the single pion effective 
action,  
and can potentially explain the oscillatory behavior seen in extracted energies 
as a function of the fit window.

Lastly, 
our computations have been carried out at a single value for the pion mass,
and with vanishing electric charges for the sea quarks.  
Further studies are required to extract physical results. 
Nevertheless, 
we have demonstrated new techniques for background field simulations
which warrant such further studies. 
We additionally intend to study spin-$1/2$ baryons using generalizations
of the methods presented here.

\bigskip
\noindent
{\bf Acknowledgements }

Lattice QCD computations were performed on clusters at Jefferson Laboratory
using time awarded through the USQCD collaboration, 
and made possible by the SciDAC Initiative. 
Support for this work was provided by the U.S.~Dept.~of Energy, 
under Grant Nos.~DE-AC05-06OR-23177 (W.D.), 
DE-FG02-04ER-41302 (W.D.), 
DE-FG02-93ER-40762 (B.C.T.), 
and DE-FG02-07ER-41527 (A.W.-L.).
Additional aspects of background field calculations reviewed in this talk were done in collaboration with J.~Hu, 
and F.-J.~Jiang.



\def\Discussion{
\setlength{\parskip}{0.3cm}\setlength{\parindent}{0.0cm}
     \bigskip\bigskip      {\Large {\bf Discussion}} \bigskip}
\def\speaker#1{{\bf #1:}\ }
\def\endDiscussion{}

\Discussion

\speaker{J. Rafelski (University of Arizona)}  
Throughout you have referred to the neutral pion. 
Is this correct? 
What about isospin?

\speaker{Tiburzi}
In background electromagnetic fields, 
isospin symmetry can no longer be used to simplify the calculation of neutral pion correlation functions. 
We have only calculated the connected part of the correlation function, 
and so 
``neutral pion'' 
should be thought of as a convenient name rather than the physical propagating state. 
The calculation of the disconnected part is left for future investigation.

\endDiscussion

 

\begin{thebibliography}{99}
\expandafter\ifx\csname natexlab\endcsname\relax\def\natexlab#1{#1}\fi
\expandafter\ifx\csname bibnamefont\endcsname\relax
  \def\bibnamefont#1{#1}\fi
\expandafter\ifx\csname bibfnamefont\endcsname\relax
  \def\bibfnamefont#1{#1}\fi
\expandafter\ifx\csname citenamefont\endcsname\relax
  \def\citenamefont#1{#1}\fi
\expandafter\ifx\csname url\endcsname\relax
  \def\url#1{\texttt{#1}}\fi
\expandafter\ifx\csname urlprefix\endcsname\relax\def\urlprefix{URL }\fi
\providecommand{\bibinfo}[2]{#2}
\providecommand{\eprint}[2][]{\url{#2}}


\bibitem{Detmold:2009dx}
\bibinfo{author}{\bibfnamefont{W.}~\bibnamefont{Detmold}},
  \bibinfo{author}{\bibfnamefont{B.~C.} \bibnamefont{Tiburzi}},
  \bibnamefont{and}
  \bibinfo{author}{\bibfnamefont{A.}~\bibnamefont{Walker-Loud}},
  \bibinfo{journal}{Phys. Rev.} \textbf{\bibinfo{volume}{D79}},
  \bibinfo{pages}{094505} (\bibinfo{year}{2009}), \eprint{0904.1586}.
  
\bibitem{Edwards:2008ja}
\bibinfo{author}{\bibfnamefont{R.~G.} \bibnamefont{Edwards}},
  \bibinfo{author}{\bibfnamefont{B.}~\bibnamefont{Joo}}, \bibnamefont{and}
  \bibinfo{author}{\bibfnamefont{H.-W.} \bibnamefont{Lin}},
  \bibinfo{journal}{Phys. Rev.} \textbf{\bibinfo{volume}{D78}},
  \bibinfo{pages}{054501} (\bibinfo{year}{2008}), \eprint{0803.3960}.

\bibitem{Lin:2008pr}
\bibinfo{author}{\bibfnamefont{H.-W.} \bibnamefont{Lin}} \bibnamefont{et~al.}
  (\bibinfo{collaboration}{Hadron Spectrum}), \bibinfo{journal}{Phys. Rev.}
  \textbf{\bibinfo{volume}{D79}}, \bibinfo{pages}{034502}
  (\bibinfo{year}{2009}), \eprint{0810.3588}.
  
\bibitem{'tHooft:1979uj}
\bibinfo{author}{\bibfnamefont{G.}~\bibnamefont{'t~Hooft}},
  \bibinfo{journal}{Nucl. Phys.} \textbf{\bibinfo{volume}{B153}},
  \bibinfo{pages}{141} (\bibinfo{year}{1979}).
  
\bibitem{Smit:1986fn}
\bibinfo{author}{\bibfnamefont{J.}~\bibnamefont{Smit}} \bibnamefont{and}
  \bibinfo{author}{\bibfnamefont{J.~C.} \bibnamefont{Vink}},
  \bibinfo{journal}{Nucl. Phys.} \textbf{\bibinfo{volume}{B286}},
  \bibinfo{pages}{485} (\bibinfo{year}{1987}). 

\bibitem{Detmold:2008xk}
\bibinfo{author}{\bibfnamefont{W.}~\bibnamefont{Detmold}},
  \bibinfo{author}{\bibfnamefont{B.~C.} \bibnamefont{Tiburzi}},
  \bibnamefont{and}
  \bibinfo{author}{\bibfnamefont{A.}~\bibnamefont{Walker-Loud}}
  (\bibinfo{year}{2008}), \eprint{0809.0721}.

\bibitem{Tiburzi:2008ma}
\bibinfo{author}{\bibfnamefont{B.~C.} \bibnamefont{Tiburzi}},
  \bibinfo{journal}{Nucl. Phys.} \textbf{\bibinfo{volume}{A814}},
  \bibinfo{pages}{74} (\bibinfo{year}{2008}), \eprint{0808.3965}.

\bibitem{Detmold:2006vu}
\bibinfo{author}{\bibfnamefont{W.}~\bibnamefont{Detmold}},
  \bibinfo{author}{\bibfnamefont{B.~C.} \bibnamefont{Tiburzi}},
  \bibnamefont{and}
  \bibinfo{author}{\bibfnamefont{A.}~\bibnamefont{Walker-Loud}},
  \bibinfo{journal}{Phys. Rev.} \textbf{\bibinfo{volume}{D73}},
  \bibinfo{pages}{114505} (\bibinfo{year}{2006}), \eprint{hep-lat/0603026}.

\bibitem{Schwinger:1951nm}
\bibinfo{author}{\bibfnamefont{J.~S.} \bibnamefont{Schwinger}},
  \bibinfo{journal}{Phys. Rev.} \textbf{\bibinfo{volume}{82}},
  \bibinfo{pages}{664} (\bibinfo{year}{1951}).

\bibitem{Hu:2007ts}
\bibinfo{author}{\bibfnamefont{J.}~\bibnamefont{Hu}},
  \bibinfo{author}{\bibfnamefont{F.-J.} \bibnamefont{Jiang}}, \bibnamefont{and}
  \bibinfo{author}{\bibfnamefont{B.~C.} \bibnamefont{Tiburzi}},
  \bibinfo{journal}{Phys. Rev.} \textbf{\bibinfo{volume}{D77}},
  \bibinfo{pages}{014502} (\bibinfo{year}{2008}), \eprint{0709.1955}.

\bibitem{Tiburzi:2007ep}
\bibinfo{author}{\bibfnamefont{B.~C.} \bibnamefont{Tiburzi}},
  \bibinfo{journal}{Phys. Rev.} \textbf{\bibinfo{volume}{D77}},
  \bibinfo{pages}{014510} (\bibinfo{year}{2008}),
  \eprint{0710.3577}.

\bibitem{Hu:2007eb}
\bibinfo{author}{\bibfnamefont{J.}~\bibnamefont{Hu}},
  \bibinfo{author}{\bibfnamefont{F.-J.} \bibnamefont{Jiang}}, \bibnamefont{and}
  \bibinfo{author}{\bibfnamefont{B.~C.} \bibnamefont{Tiburzi}},
  \bibinfo{journal}{Phys. Lett.} \textbf{\bibinfo{volume}{B653}},
  \bibinfo{pages}{350} (\bibinfo{year}{2007}), \eprint{0706.3408}.

\bibitem{Tiburzi:2008pa}
\bibinfo{author}{\bibfnamefont{B.~C.} \bibnamefont{Tiburzi}},
  \bibinfo{journal}{Phys. Lett.} \textbf{\bibinfo{volume}{B674}},
  \bibinfo{pages}{336} (\bibinfo{year}{2009}),
  \eprint{0809.1886}.










\end{thebibliography}
\end{document}